\documentclass[aps,prl,preprint,tightenlines,superscriptaddress,showpacs,byrevtex]{revtex4}
\usepackage{graphicx} 
\usepackage{dcolumn}  
\usepackage{bm}

\graphicspath{{ps}}

\renewcommand{\arraystretch}{1.1}

\begin{document}


\preprint{\vbox{ 
		 \hbox{BELLE-PREPRINT-2003-12}
		 \hbox{KEK-PREPRINT-2003-44}
}}

\title{ \quad\\[0.5cm]  Search for $B^0\to\ell^+\ell^-$ at Belle}
\affiliation{Budker Institute of Nuclear Physics, Novosibirsk}
\affiliation{Chiba University, Chiba}
\affiliation{University of Cincinnati, Cincinnati, Ohio 45221}
\affiliation{Gyeongsang National University, Chinju}
\affiliation{University of Hawaii, Honolulu, Hawaii 96822}
\affiliation{High Energy Accelerator Research Organization (KEK), Tsukuba}
\affiliation{Hiroshima Institute of Technology, Hiroshima}
\affiliation{Institute of High Energy Physics, Chinese Academy of Sciences, Beijing}
\affiliation{Institute of High Energy Physics, Vienna}
\affiliation{Institute for Theoretical and Experimental Physics, Moscow}
\affiliation{J. Stefan Institute, Ljubljana}
\affiliation{Kanagawa University, Yokohama}
\affiliation{Korea University, Seoul}
\affiliation{Kyungpook National University, Taegu}
\affiliation{Institut de Physique des Hautes \'Energies, Universit\'e de Lausanne, Lausanne}
\affiliation{University of Ljubljana, Ljubljana}
\affiliation{University of Maribor, Maribor}
\affiliation{University of Melbourne, Victoria}
\affiliation{Nagoya University, Nagoya}
\affiliation{Nara Women's University, Nara}
\affiliation{National Lien-Ho Institute of Technology, Miao Li}
\affiliation{Department of Physics, National Taiwan University, Taipei}
\affiliation{H. Niewodniczanski Institute of Nuclear Physics, Krakow}
\affiliation{Nihon Dental College, Niigata}
\affiliation{Niigata University, Niigata}
\affiliation{Osaka City University, Osaka}
\affiliation{Osaka University, Osaka}
\affiliation{Panjab University, Chandigarh}
\affiliation{Peking University, Beijing}
\affiliation{Princeton University, Princeton, New Jersey 08545}
\affiliation{Saga University, Saga}
\affiliation{University of Science and Technology of China, Hefei}
\affiliation{Seoul National University, Seoul}
\affiliation{Sungkyunkwan University, Suwon}
\affiliation{University of Sydney, Sydney NSW}
\affiliation{Tata Institute of Fundamental Research, Bombay}
\affiliation{Toho University, Funabashi}
\affiliation{Tohoku Gakuin University, Tagajo}
\affiliation{Tohoku University, Sendai}
\affiliation{Department of Physics, University of Tokyo, Tokyo}
\affiliation{Tokyo Institute of Technology, Tokyo}
\affiliation{Tokyo Metropolitan University, Tokyo}
\affiliation{Tokyo University of Agriculture and Technology, Tokyo}
\affiliation{Toyama National College of Maritime Technology, Toyama}
\affiliation{University of Tsukuba, Tsukuba}
\affiliation{Utkal University, Bhubaneswer}
\affiliation{Virginia Polytechnic Institute and State University, Blacksburg, Virginia 24061}
\affiliation{Yokkaichi University, Yokkaichi}
\affiliation{Yonsei University, Seoul}
  \author{M.-C.~Chang}\affiliation{Department of Physics, National Taiwan University, Taipei} 
  \author{P.~Chang}\affiliation{Department of Physics, National Taiwan University, Taipei} 
  \author{K.~Abe}\affiliation{High Energy Accelerator Research Organization (KEK), Tsukuba} 
  \author{T.~Abe}\affiliation{High Energy Accelerator Research Organization (KEK), Tsukuba} 
  \author{I.~Adachi}\affiliation{High Energy Accelerator Research Organization (KEK), Tsukuba} 
  \author{H.~Aihara}\affiliation{Department of Physics, University of Tokyo, Tokyo} 
  \author{M.~Akatsu}\affiliation{Nagoya University, Nagoya} 
  \author{Y.~Asano}\affiliation{University of Tsukuba, Tsukuba} 
  \author{T.~Aso}\affiliation{Toyama National College of Maritime Technology, Toyama} 
  \author{V.~Aulchenko}\affiliation{Budker Institute of Nuclear Physics, Novosibirsk} 
% ...
\collaboration{The Belle Collaboration}

\begin{abstract}
We report the results of a search for the decay
$B^0\to e^+ e^-$, $\mu^+ \mu^-$ and $e^{\pm} \mu^{\mp}$  
based on an
analysis of 78 $\rm fb^{-1}$ of data collected by the Belle detector at 
KEKB. No candidate events have been found.
Upper limits on the branching fractions are calculated at the 90\%
confidence level:
$\mathcal {B}$($B^0$ $\to$ $e^+$ $e^-$) $<$ 1.9 $\times$ $10^{-7}$, 
$\mathcal {B}$($B^0$ $\to$ $\mu^+$ $\mu^-$) $<$ 1.6 $\times$ $10^{-7}$, 
and 
$\mathcal {B}$($B^0$ $\to$ $e^{\pm}$ $\mu^{\mp}$) $<$ 1.7 
$\times$ $10^{-7}$.
A limit on the Pati-Salam leptoquark mass $M_{\rm LQ}$ $>$ 46 TeV/$c^2$ is 
obtained at the 90\% confidence level.
\end{abstract}

\maketitle

\tighten

{\renewcommand{\thefootnote}{\fnsymbol{footnote}}}
\setcounter{footnote}{0}


The $B$ meson decays $B^0\to e^+ e^-$ and $B^0\to \mu^+ \mu^-$
can proceed at a low rate through the
Flavor-Changing Neutral Current (FCNC) process. 
The Standard Model (SM) predictions for the branching fractions
are  $(2.34 \pm 0.33) \times 10^{-15}$ and $(1.00 \pm 0.14) \times 10^{-10}$
for 
$B^0 \to e^+ e^-$ and $B^0 \to \mu^+ \mu^-$ decays, respectively~\cite{buras}. 
These branching fractions could be two
orders of magnitude larger in models with two Higgs doublets and
$Z$-mediated FCNC~\cite{theorist}.
The decays $B^0$ $\to$ $e^{\pm}$ $\mu^{\mp}$ 
are essentially forbidden in the SM by 
lepton number conservation,
apart from a vanishingly small probability permitted by neutrino oscillation.
They can occur, however, in the Pati-Salam 
Model~\cite{pati} or supersymmetric (SUSY) models~\cite{susy}.
The charge conjugate modes are implicitly included throughout this letter.

In this letter we report 
the results of
a search for the decays $B^0 \to e^+e^-$, $\mu^+\mu^-$ and $e^\pm \mu^\mp$
(collectively denoted as $B^0\to\ell^+\ell^-$) using 78 $\rm fb^{-1}$ of data,
corresponding to 85 million $B\overline{B}$ pairs,
 collected with the Belle detector at
KEKB~\cite{KEKB}, an asymmetric $e^+$$e^-$ collider operating at 
the $\Upsilon(4S)$ resonance.
The Belle detector is a large-solid-angle magnetic
spectrometer that
consists of a three-layer silicon vertex detector (SVD),
a 50-layer central drift chamber (CDC), an array of
aerogel threshold Cherenkov counters (ACC), time-of-flight
scintillation counters (TOF), and an array of CsI(Tl) crystals
(ECL)  located inside
a superconducting solenoid coil that provides a 1.5 T
magnetic field.  An iron flux-return located outside of
the coil is instrumented to 
detect $K_L^0$ mesons and to identify muons (KLM).
The detector is described in detail elsewhere~\cite{Belle}.  
    
For each primary charged track, the impact parameter
relative to the run-by-run interaction point is required to be within
2~cm along the $z$ axis (aligned with the positron beam) and within
0.08~cm in the plane transverse to this axis, and the transverse
momentum must exceed 100 MeV/$c$.  At least five charged tracks 
including non-primary tracks are required.
For each event the visible energy in the
detector is required to be greater than 0.6 GeV. 
Monte Carlo (MC) 
studies show that the radiative
Bhabha events (100\%), as well as two-photon (94\%) and $\tau^+\tau^-$ (85\%) events,
are rejected by these criteria.
The two-photon MC studies use the
generator AAFH~\cite{CPC}, which includes all tree-level diagrams
to fourth order in $\alpha_{\rm QED}$.

Electron tracks are distinguished from hadron tracks by examining the 
amount of energy and the transverse shower profile of the associated ECL cluster
as well as light yield in the ACC. The $dE/dx$ measurements in the CDC is also taken into account. 
Muon tracks are distinguished from hadron tracks by their
range and  transverse scattering in the KLM.
We select electrons and muons by requiring that the normalized
likelihood~\cite{nim-eid, nim-muid} that the track is a lepton 
rather than a hadron exceeds 0.9. The electron and the muon 
detection efficiencies are 88\% and 89\% which are studied via 
$J/\psi \to e^+ e^-$ and $J/\psi \to \mu^+ \mu^-$ in the same track momentum range, 1.5 GeV/$c$ - 4.5 GeV/$c$. 
The corresponding misidentification rates from kaon tracks and pion tracks are 
1.7\% ($K \to \mu$), 1.0\% ($\pi \to \mu$), $<$ 0.005\% ($K \to e$), and $<$ 0.1\% ($\pi \to e$)
which are studied using kaons and pions from the decay chain 
$D^* \to D^0 \pi^- \to (K^- \pi^+) \pi^-$ in the same track momentum range, 
1.5 GeV/$c$ - 4.5 GeV/$c$. 

$B$ meson candidates are formed from pairs of
oppositely charged leptons, not necessarily of the same flavor,
and selected using two kinematic variables 
defined in the $\Upsilon(4S)$ center-of-mass (CM) frame:
the beam-energy 
constrained mass, $M_{\rm bc} = $ $\sqrt{E^{2}_{\rm beam} - p^{2}_{B}}$, 
and the energy difference, $\Delta E = $ $E_B - E_{\rm beam}$, 
where $p_B$ and $E_B$ are the momentum and energy,
respectively, of the $B$ candidate and $E_{\rm beam}$ is the
beam energy.
The signal region is defined as 5.27 GeV/$c^2$ $<$ $M_{\rm bc}$ $<$ 
5.29 GeV/$c^2$ and $|\Delta E|$ $<$ 0.05 GeV for all 
channels. The sideband region, used for the background
determination, is defined as 5.20 GeV/$c^2$ $<$ 
$M_{\rm bc}$ $<$ 5.26 GeV/$c^2$ and $|\Delta E|$ $<$ 0.30 GeV. 

\begin{table}[htb]
\caption{The number of expected events are normalized
to the integrated luminosity of the data sample.
The results show that the backgrounds
in MC, off-resonance data
and on-resonance data are in good agreement after
the $N\ge5$ cut.}
\label{tab:background}
\begin{tabular}{lccc}
\hline \hline
Mode& $N^{\rm trk=3, 4}$ & $N^{\rm trk\ge 5}$ & $N^{\rm trk\ge 5}_{\cal LR \rm >0.9}$\\
\hline
$\tau^+\tau^-$ & $14 \pm 4$ & $2.4 \pm 1.7$ & $0$ \\
two photon & $250 \pm 109$ & $16 \pm 5$ & $0$\\
$q\overline{q}$ ($q=u,d,s$)& $6 \pm 1$ &$82 \pm 7$ & $1.3 \pm 0.9$ \\
$q\overline{q}$ ($q=c$)& $54 \pm 6$ &$396 \pm 16$ & $3.8 \pm 1.5$ \\
off-resonance data & $396 \pm 46$ & $431 \pm 46$ & $10 \pm 7$ \\
data & 441 & 473 & 12 \\
\hline \hline
\end{tabular}
\end{table}

\begin{figure}[!htb]
\begin{center}
\resizebox*{1.66in}{1.7in}{\includegraphics{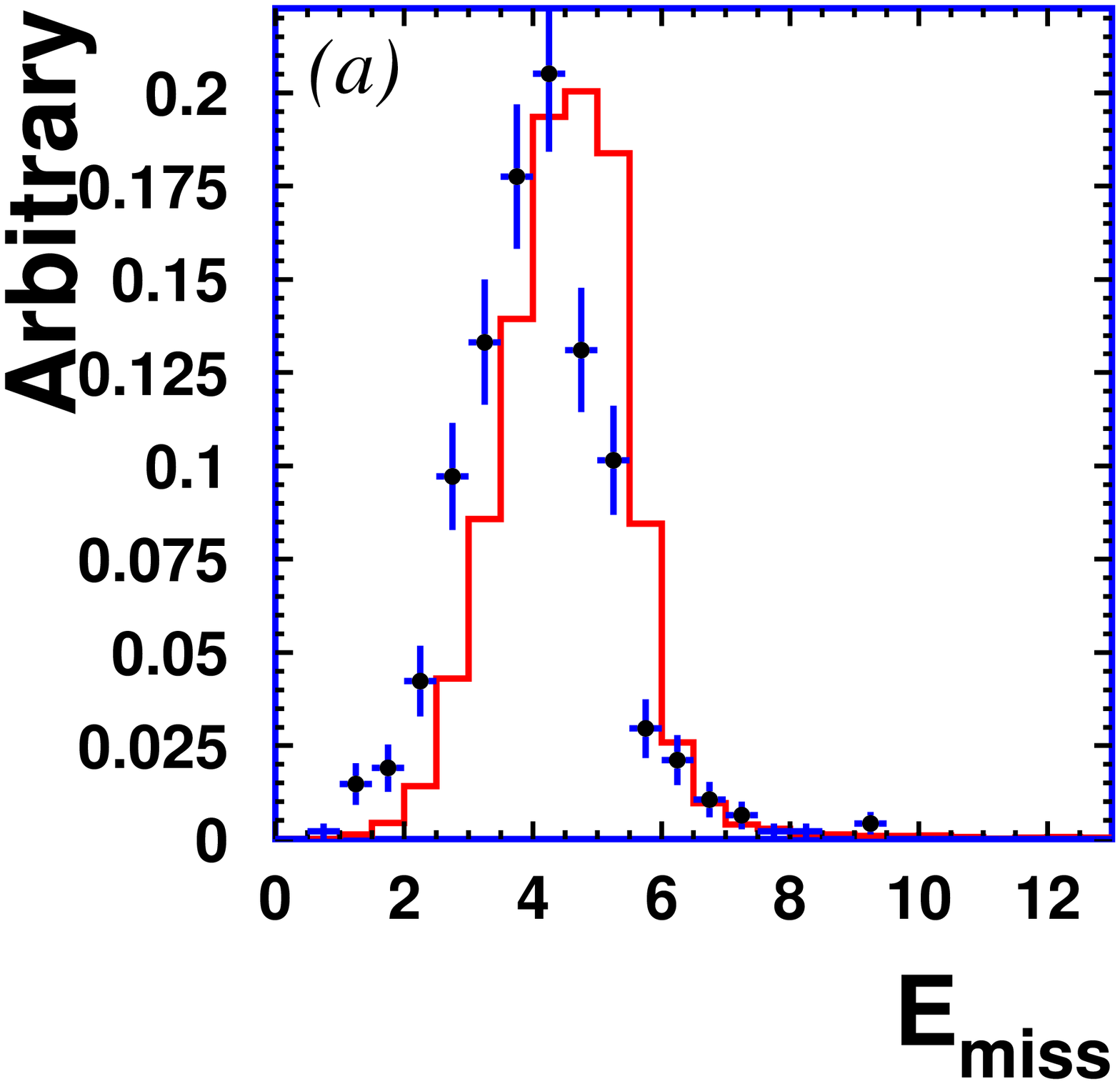}}
\resizebox*{1.66in}{1.7in}{\includegraphics{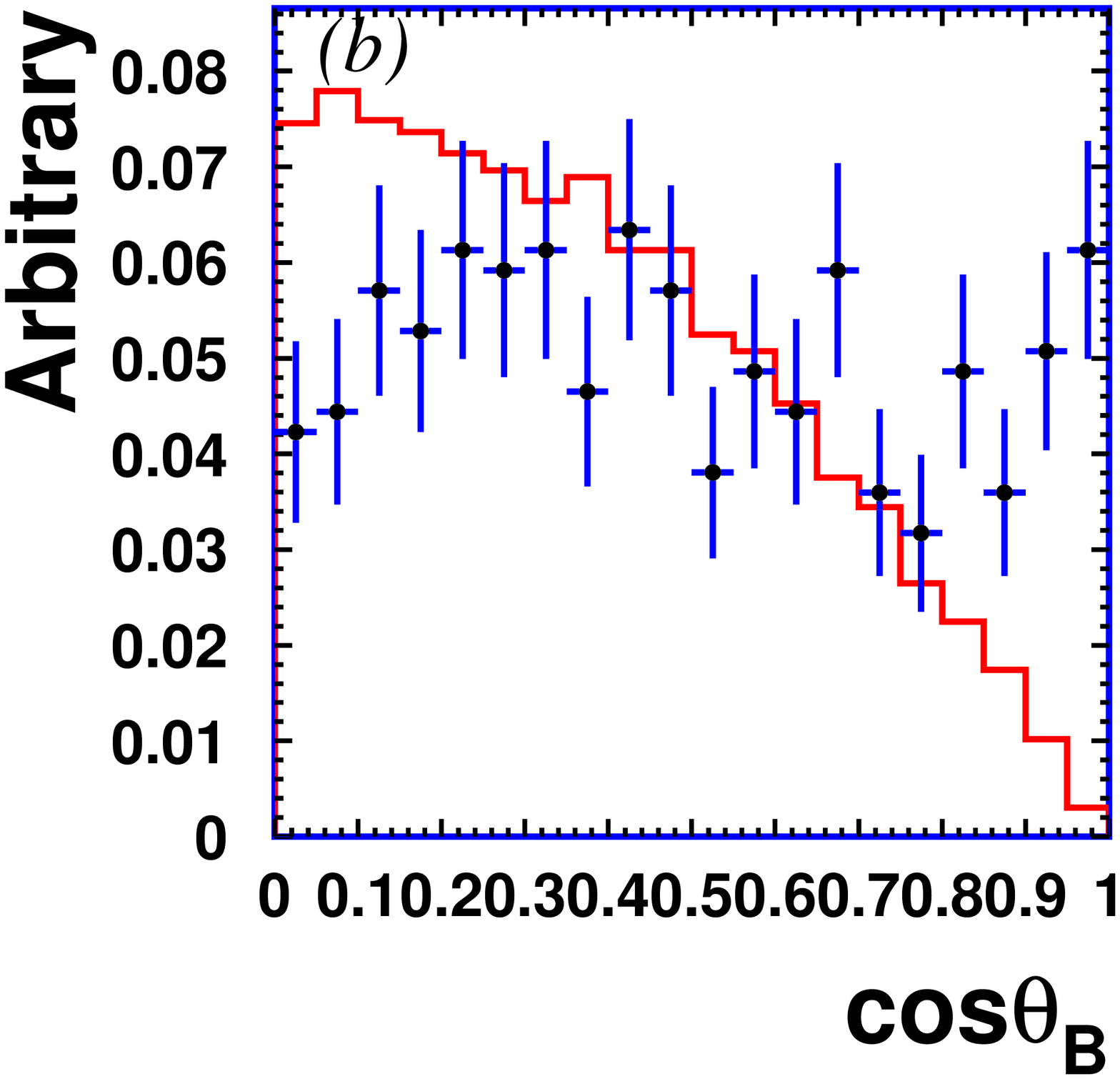}}
\resizebox*{1.66in}{1.7in}{\includegraphics{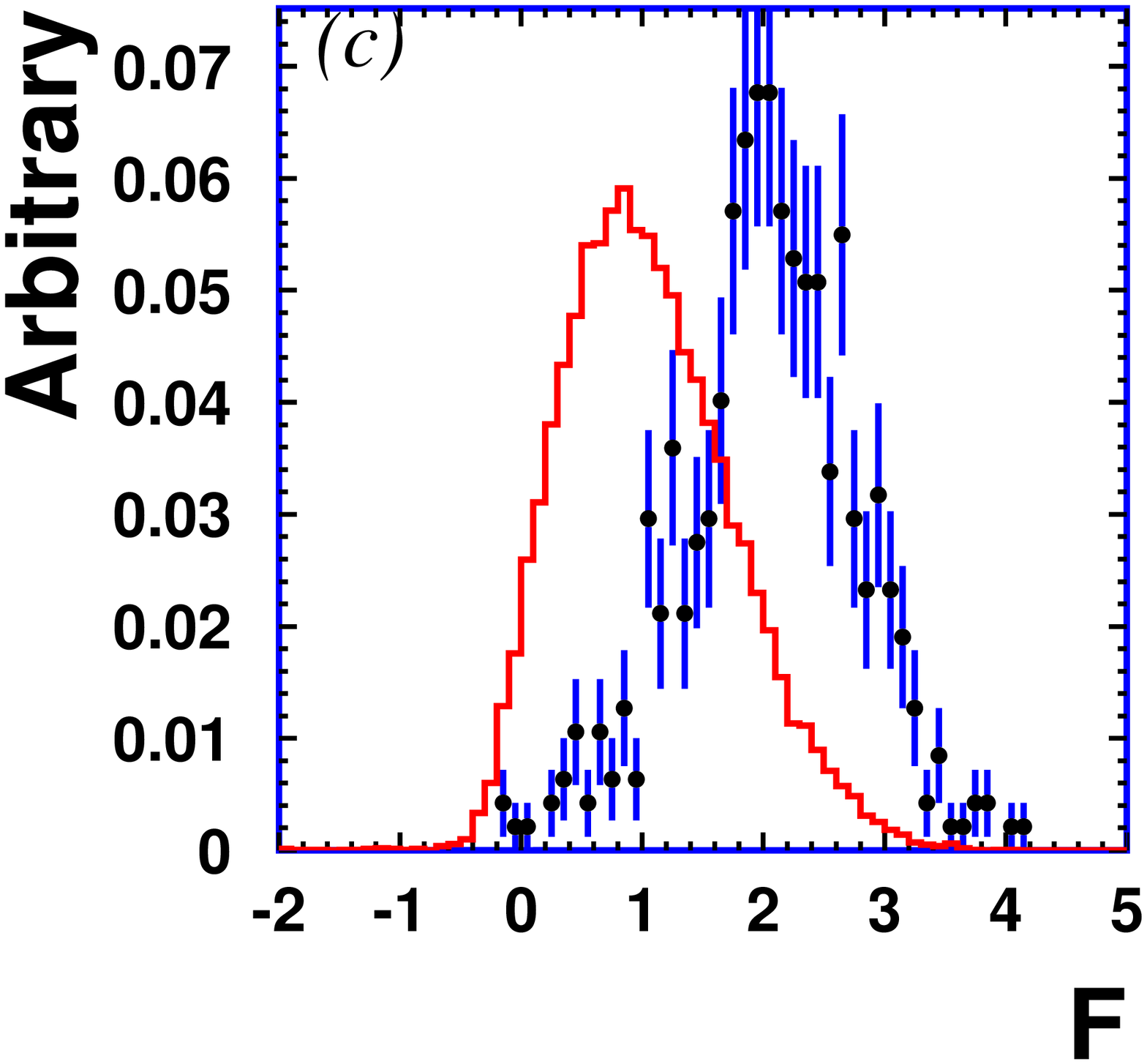}}
\resizebox*{1.66in}{1.7in}{\includegraphics{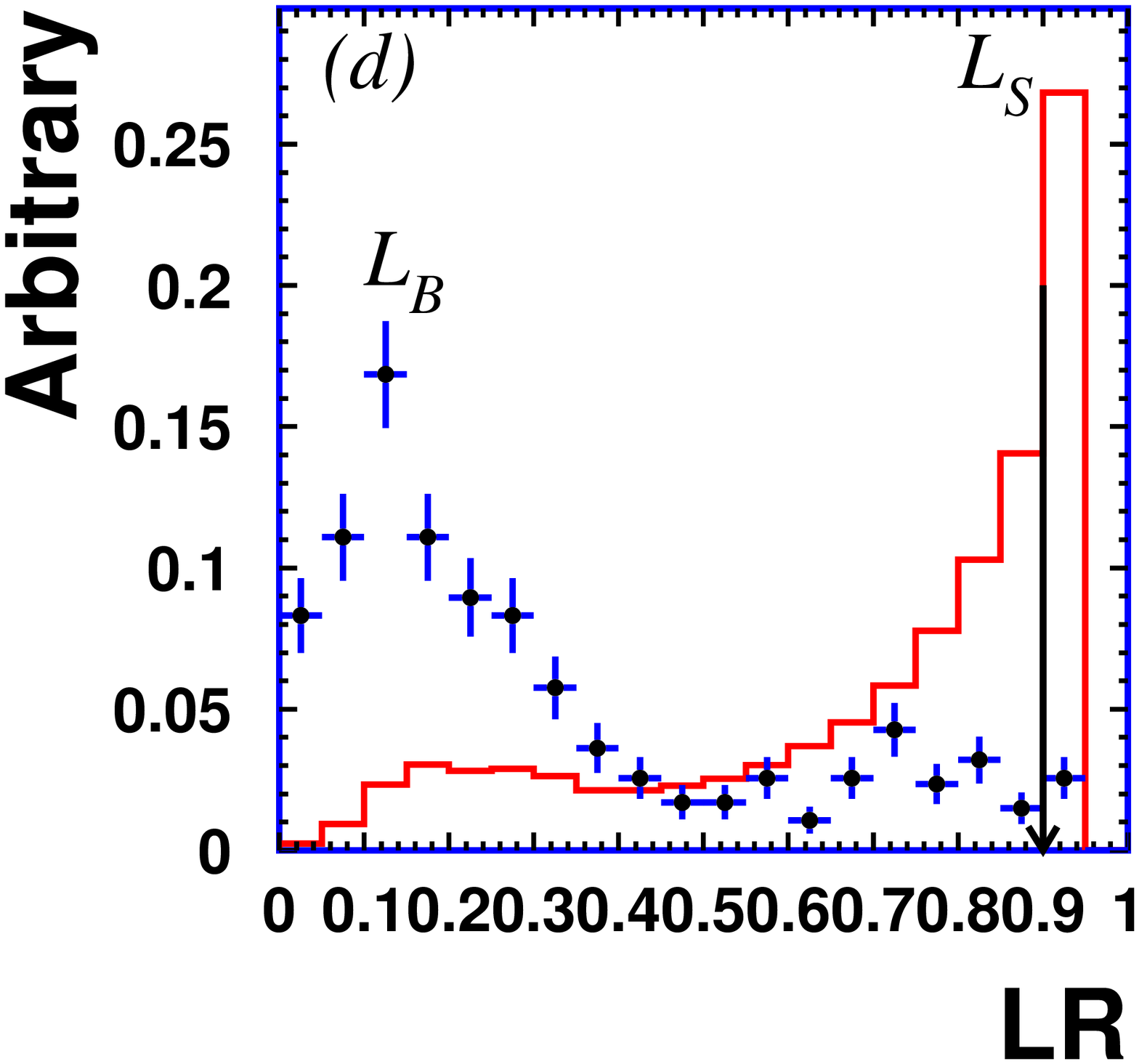}}
\end{center}
\caption{Open histograms are from signal MC and
the points with error bars are from sideband data.
The arrow in the (d) $\cal LR$ plot indicates the optimized selection.}
\label{fig:lr}
\end{figure}

The decays $B^0 \to K^+\pi^-$ and $B^0 \to \pi^+\pi^-$
do not contribute measurably to the background because
of their small branching fractions, the small lepton
fake rates, and, for the $K^+\pi^-$ mode, the shift of
the $B$ candidate out of the $\Delta E$ signal region
due to the use of lepton masses for the tracks.  
Background from other $B$ decays is also found to be negligible.
Backgrounds from continuum $e^+ e^- \to q\overline{q}$
($q=u,d,s,c$) production are suppressed using seven variables
that characterize the event topology: five modified Fox-Wolfram
moments~\cite{sfw}, $S_{\perp}$~\cite{sperp}, and the cosine of
the polar angle $\theta_{\rm thr}$ of the $B$ candidate's thrust
axis~\cite{fisher}. Events $e^+ e^- \to c\overline{c}$ with two semileptonic
charm decays produce $B^0 \to \ell^+ \ell^-$ candidates which are
not suppressed by lepton identification cuts; without special
treatment, these events would be the dominant source of
background. We use four additional variables to suppress these
events: the missing energy $E_{\rm miss}$  and momentum
$p_{\rm miss}$, the cosine of the angle $\theta_{\rm pl}$ between
the missing momentum and the $B$ candidate's thrust axis, the
cosine of the angle $\theta_{\rm pb}$ between the missing momentum
and the thrust axis of the remaining particles in the event.  All
eleven variables are combined into a Fisher discriminant~\cite{fisher}. 
We obtain probability density functions (PDFs) for
the signal and for continuum background as functions of this Fisher
discriminant ($F$) and of the cosine of the polar angle ($\cos \theta_B$) of
the $B$ candidate's flight direction.  We then obtain signal and
background likelihoods, ${\cal L}_S$ and ${\cal L}_B$, from the
product of the PDFs for $F$ and $\cos \theta_B$, and demand that
${\cal LR}={\cal L}_S/({\cal L}_S + {\cal L}_B)$ exceeds 0.9.  This
criterion was chosen to optimize the performance of the 
signal significance ($N_S/\sqrt {N_S+N_B}$) using both MC and data samples, where $N_S$ and $N_B$ are predicted signal yields
and predicted background yields, respectively. 
Assuming the branching fractions of $B^0 \to \ell^+ \ell^-$ are 10$^{-7}$ $\sim$ 10$^{-15}$, we then obtain the signal yields from the product of the total number of $B\overline{B}$ events, the signal efficiency from MC and the assumed branching fractions.
The predicted background yields are obtained by fitting the distribution of $\Delta E$ from the sideband data projection in the signal region. 
The backgrounds in the on-resonance
sideband region are compared
to MC and to data taken 50-60 MeV below
the $\Upsilon(4S)$ resonance. The results are
summarized in Table~\ref{tab:background}.
Distributions of $E_{\rm miss}$, $\cos \theta_B$, $F$ and 
${\cal LR}$ are shown in Fig.~\ref{fig:lr}.

The signal
efficiency for each mode, after application of all selection criteria, is
determined by MC simulation and itemized in Table~\ref{tab:conclude}.
The systematic error
of this efficiency arises from tracking efficiency (1\% per track), 
 electron and
muon identification efficiencies (4\% and 2\% per lepton, respectively, determined
from a study of $J/\psi$ decays), the charged-track-count
criterion (10\%, measured from the uncertainty in the efficiency using a $B^0 \to K^+ \pi^-$ control sample),
and the likelihood  ratio cut (4\%, based on a study of $B^- \to D^0 \pi^- \to
(K^- \pi^+) \pi^-$).
The total systematic uncertainty is obtained by first summing the
correlated errors linearly and then adding the non-correlated
errors in quadrature.

Since no signal events are found for any mode, we
determine 90\% confidence level upper limits on the branching
fractions by an extension of the Feldman-Cousins method,
which takes 
systematic uncertainties into account~\cite{fc1}.
The $q\overline{q}$ events are the main background after all the event selections. 
In order to estimate the number of expected background events in the
signal box, a $q\overline{q}$ MC sample (corresponding to an integrated luminosity 1.6 times greater than that of the data) is used
to fit the shape of the background
distribution in $\Delta E$ and $M_{\rm bc}$ in the sideband region. These
distributions are scaled to the number of observed data events in the
sideband region; the number of expected background events is
then obtained from the signal-to-sideband ratio.  
The dominant systematic uncertainty on the background estimation
comes from the statistical error of the number of sideband data events.      
The results are listed in Table~\ref{tab:conclude}.

\begin{table*}[htb]
\caption{Summary of the  $B^0\to \ell^+ \ell^-$ search, where 
$\epsilon$ is the reconstruction efficiency,
$N_{\rm obs}$ is the measured number of events in the signal region,
$N^{\rm bg}_{\rm exp}$ is the expected background in the signal region
and BF is the 90\% confidence level upper limit for the branching fractions.}
\label{tab:conclude}
\newcommand{\m}{\hphantom{$-$}}
\newcommand{\cc}[1]{\multicolumn{1}{c}{#1}}
\renewcommand{\tabcolsep}{1.2pc} 
\renewcommand{\arraystretch}{1.2} 
\begin{tabular}{@{}lcccl}
\hline \hline
Mode& $\epsilon$[\%] &$N_{\rm obs}$  &$N^{\rm bg}_{\rm exp}$ & BF($10^{-7}$) \\
\hline
$B^0 \to e^+ e^-$ & $14.3 \pm 2.0$ & 0 & $0.30 \pm 0.12$   & $<1.9$ \\
$B^0 \to \mu^+ \mu^-$ & $16.9 \pm 2.0$ & 0 & $0.19 \pm 0.10$  & $<1.6$ \\
$B^0 \to e^{\pm} \mu^{\mp}$ & $15.8 \pm 1.9$ & 0  & $0.22 \pm 0.10$ & $<1.7$ \\
\hline \hline
\end{tabular}
\end{table*}

These upper limits can be used to constrain certain
extensions of the Standard Model.  For example, the Pati-Salam
model~\cite{LQ2} predicts the existence of leptoquarks---heavy
spin-1 gauge bosons that carry both color and lepton quantum 
number---that mediate the decays $B^0 \to e^{\pm} \mu^{\mp}$~\cite{LQ1, LQ2}.
The branching ratio is related to the leptoquark mass $M_{\rm LQ}$ 
by~\cite{LQ2} 
$$\Gamma (B^0 \to e^{\pm} \mu^{\mp}) = \pi \alpha_s^2(M_{LQ}) 
\frac {1}{M_{LQ}^4} F_{B^0}^2 m_{B^0}^3 R^2,$$
where
$$R = \frac {m_{B^0}}{m_b} 
\left (\frac {\alpha_s (M_{LQ})}{\alpha_s (m_t)}\right)^{-4/7} 
\left (\frac {\alpha_s (m_t)}{\alpha_s (m_b)}\right)^{-12/23}\, .$$

We follow the method in Refs.~\cite{pub3, pub4} and use
$F_B$ = 175 $\pm$ 30 MeV for the $B^0$ decay constant,
$m_{B^0}$ = 5279.3 $\pm$ 0.7 MeV/$c^2$~\cite{pub6} for the $B^0$ meson mass,
$m_b$ = 4.3 $\pm$ 0.2 GeV/$c^2$~\cite{pub3} for the $b$-quark mass
and $m_t$ = 176.0 $\pm$ 6.5 GeV/$c^2$~\cite{pub6} for the $t$-quark mass.
We use $\alpha_s$ ($M_Z$) = 0.117~\cite{pub6} which is evolved to 
$M_{\rm LQ}$ using the Marciano approximation~\cite{LQ5}, assuming
no other colored particles exist between $m_t$ and $M_{\rm LQ}$.
We obtain a limit of $M_{\rm LQ} > 46\,{\rm TeV}/c^2$
on the mass of the Pati-Salam leptoquark at the 90\% confidence level.
This bound holds in the case where
the $\tau$ lepton is associated with the strange and charm quarks
within the Pati-Salam theory. Previous lower limits for this case
from $B^+ \to \ell^+\nu$~\cite{LQ1,LQ3} and $B^0 \to \mu e$~\cite{pub4,pub3} are weaker than the bound presented
here; the limit from lepton universality in $\pi^+ \to \ell^+ \nu$ is
stronger~\cite{LQ1,LQ2}.               
	
In conclusion, we find no evidence for $B^0 \to
e^+ e^-$, $\mu^+ \mu^-$, and $e^{\pm} \mu^{\mp}$ decays in
85 million $B\overline{B}$ events, and set
upper limits on the branching fractions of $1.9 \times 10^{-7}$,
$1.6 \times 10^{-7}$, and $1.7 \times 10^{-7}$, respectively, at the
90\% confidence level.  We use the latter upper limit to set a
lower bound on the mass of the Pati-Salam leptoquark of 46 TeV/$c^2$.

We wish to thank the KEKB accelerator group for the excellent
operation of the KEKB accelerator.
We acknowledge support from the Ministry of Education,
Culture, Sports, Science, and Technology of Japan
and the Japan Society for the Promotion of Science;
the Australian Research Council
and the Australian Department of Education, Science and Training;
the National Science Foundation of China under contract No.~10175071;
the Department of Science and Technology of India;
the BK21 program of the Ministry of Education of Korea
and the CHEP SRC program of the Korea Science and Engineering Foundation;
the Polish State Committee for Scientific Research
under contract No.~2P03B 01324;
the Ministry of Science and Technology of the Russian Federation;
the Ministry of Education, Science and Sport of the Republic of Slovenia;
the National Science Council and the Ministry of Education of Taiwan;
and the U.S.\ Department of Energy.


\end{document}